\documentstyle[prl,aps,12pt]{revtex}
\input{epsf.sty}

\def\bp{{\bf p}}
\def\bk{{\bf k}}
\def\bD{{\bf D}}
\def\b\mu{{\bf \mu}}

\begin{document}

\draft

\title{The Fermion Boson Interaction Within the Linear Sigma Model \\
at Finite Temperature}
\author{H.C. de Godoy Caldas\thanks{e-mail: hcaldas@funrei.br}}

\address{Departamento de Ci\^{e}ncias Naturais, DCNAT \\
Funda\c{c}\~ao de Ensino Superior de S\~{a}o Jo\~{a}o del Rei, FUNREI,\\ 
Pra\c{c}a Dom Helv\'ecio, 74, CEP:36300-000, S\~{a}o Jo\~{a}o del Rei, MG, Brazil} 

\date{January, 2001}

\maketitle                          

\begin{abstract}
We reinvestigate the interaction of massless fermions with massless bosons at finite temperature. Specifically, we calculate the self-energy of massless fermions 
due the interaction with massless bosons at high temperature, which is the region where thermal effects are maximal. The calculations are concentrated in the limit of vanishing fermion three momentum and after considering the 
effective fermion and boson dressed masses, we obtain the damping rate of the fermion up to order $g^3$. It is shown that in the limit $k_0 \ll T$ the fermion acquire a thermal mass of order $gT$ and the leading term of the fermion damping rate is of order $g^2T+g^3T$.

\end{abstract}

\vspace{1cm}


\newpage
 
\section{Introduction} 

The fermion dispersion relation to one-loop order at high temperature has 
been studied several times in the literature \cite{Weldon1,Kapusta1,Braaten}. The 
dispersion relation of a fermion, which gives its energy $\omega$ as it propagate through the medium as a function of its momentum $k$, is very important in different kinds of physical situations \cite{Letessier,Braaten1,Thoma1} 
and has been focused on various approximations and limits \cite{Daniel1,Weldon2}. During the last few years, there has been some controversy if the damping rate is gauge dependent or not and other problems \cite{Rebhan1}. It has been proposed that a proper resummation cures such problems found in most one-loop calculations \cite{Rebhan2}. In fact, if one wants consistent results of damping rates at positive temperatures, it is crucial the use of methods like the well known hard thermal loop (HTL) resummation of Braaten and Pisarski \cite{Braaten0} which employs resumed propagators and vertices and keep the correct order of the coupling constant. Whith this method, Braaten and Pisarski has given in ref. \cite{Braaten2} a formal proof that resummation produces gauge-invariant results for the damping rates of quarks and gluons. In this article we reinvestigate the self-energy of massless fermions interacting with massless bosons at high temperature in the framework of the linear sigma model. This is the simpler, but instructive situation of fermions interacting with scalar and pseudoscalar fields. These are similar to Yukawa couplings which has been discussed by M. Thoma in ref. \cite{Thoma}. Using the HTL, Thoma has found a kinematic restriction (not observed in the case of quarks or gluons) depending on the effective fermion and boson masses involved in the calculation of the damping rate of the fermion. More recently, in ref. \cite{Daniel1}, a fermion-scalar plasma has been considered in the context of real-time formulation and it was found that the Yukawa fermions acquire a width through the induced decay of the scalar in the medium. Here, as in \cite{Thoma}, we use imaginary time formalism and calculate the fermion dispersion relation for interacting massless bosons and fermions in the limit $k_0,k \ll T$ and compute the fermion damping rate at rest, $k=0$, for the case where the effective dressed (induced by the thermal medium) fermion and boson masses are consistently considered.

In a recent paper \cite{Caldas1} we proposed a modified self-consistent resummation (MSCR) which resums higher-order terms in a non-perturbative 
way in order to cure the problem of breakdown of the perturbative expansion 
at finite temperature. We have applied the method up to one \cite{Caldas1} and two-loop \cite{Caldas1-2} order in the perturbative expansion. 
We have shown that the MSCR, when applied to the study of the chiral fermion meson model, has the essential features which lead to the satisfaction of Goldstone's theorem and renormalization of the UV divergences, in the low 
and high temperature regions. We have explicitly shown that the scheme 
breaks down around $T_c$ i.e., in the region of intermediate temperatures, 
since quantum fluctuations are known to play a major role there. In this 
region higher-order terms in the perturbative expansion are required. 

It is well known that at high temperature the perturbative expansion can 
also be broken in theories with spontaneous symmetry breaking (SSB) or in massless field theories (like QCD) because powers of the temperature can compensate for powers of the coupling constant, even if the strength of the coupling is small 
\cite{Weinberg,Dolan}. Infrared divergence appears. When a set of infrared-divergent diagrams is 
summed up one gets an infrared-finite result. This is implemented in the 
MSCR by the recalculation of the self-energy. A comparison between the MSCR and the HTL resummation methods showing its similarities and differences, which is out of the scope of this paper, is under preparation and will be reported elsewhere \cite{Caldas2}. So, a motivation to study the fermion dispersion relation and damping rate at high temperature is the fact that the MSCR showed itself an efficient method to execute the resummation in a divergence-free way in this region. Thus, in the computation of the damping rate of the fermion, we shall use at the starting point the dressed by interaction boson mass obtained in \cite{Caldas1} rather than the zero mass parameter of the Lagrangian. In this way, this calculation is interesting since it provides another simple example for the application of the MSCR method.

This paper is organized as follows. In Section II we study the 
fermion self-energy and obtain the dispersion relation 
of the fermion with some of its interesting limits. In Section III we compute the fermion damping rate at rest. Section IV is devoted to conclusions.

\section{The Fermion Self-Energy}

We describe the fermion-bosons vertex by the interaction Lagrangian extracted from the linear sigma model \cite{Gell-Mann}

\begin{equation}
\label{LI}
{\cal L}^{int.}= -g\overline{\psi}\left[\sigma +
 i \gamma^5{\vec\pi}\cdot {\vec\tau}\right]\psi,
\end{equation}
where $\psi$, $\sigma$, and $\pi$ represent the quark, sigma and
pion fields, respectively, and $g$ is a non-dimensional positive coupling constant.

The fermion self-energy is defined by

\begin{equation}
\label{self1}
{\cal D}(\omega_{n}, \bk)^{-1}=  {\cal D}_0(\omega_{n}, \bk)^{-1}+ \Sigma(\omega_n, \bk),
\end{equation}
where ${\cal D}_0(\omega_{n}, \bp)$ is the tree-level fermion propagator, expressed as

\begin{equation}
\label{prop1}
{\cal D}_{0}( \omega_{n}, \bk)^{-1}= \gamma_{\mu} k^{\mu}- m_{\psi},
\end{equation}
and $\Sigma =\Sigma_s + \gamma^\mu \Sigma_\mu$, with $\Sigma_s$, $\Sigma_0$ 
and $\vec \Sigma$ being the contributions proportional to the unit, $\gamma^0$ and $\vec \gamma$ matrices respectively.

To one-loop order the fermion self-energy expression is given\cite{Kapusta} by

\begin{eqnarray}
\label{self2}
\Sigma(k_0,\vec{k}) = \left( \frac{\delta \ln Z_I^{2-loop}}{\delta {\cal D}_{0 \psi}} \right)_{1PI}=\\
\nonumber
- g^2T \sum_{n}\int \frac {d^3 p}{(2\pi)^3}\bD_{0 \sigma}(\omega_{n+l}, \bp+ \bk) {\cal D}_{0 \psi}(\omega_n, \bp) + \\
\nonumber
- 3g^2T \sum_{n}\int \frac {d^3 p}{(2\pi)^3}\bD_{0 \pi}(\omega_{n+l}, \bp+ \bk) {\cal D}_{0 \psi}(\omega_n, \bp),
\end{eqnarray}
since the logarithm of the two-loop interaction partition function is found to be \cite{Caldas1}:

\begin{equation}
\label{lnint}
\ln Z_I^{2-loop}= \frac {1}{2}{g^2}\int_0^{\beta} d \tau_1 d \tau_2 \int d^3 x_1 d^3 x_2 \frac {\int [d \phi]e^{S_0} [(\overline{\psi} \sigma \psi)^2 + (\overline{\psi}i \gamma^5{\vec\pi}\cdot {\vec\tau}\psi)^2]}{\int [d\phi]e^{S_0}}.
\end{equation}

In eq.(\ref{self2}) $\omega_{n}$ are the Matsubara frequencies, defined as $\omega_{n}=2n\pi T$ for bosons and $\omega_{n}=(2n+1)\pi T$ for fermions and 
$\bD_{0 \sigma,\pi}(\omega_{n}, \bp)$ is the tree-level boson propagator, expressed as

\begin{equation}
\label{propboson}
\bD_{0 \sigma,\pi}(\omega_n, \bk) ^{-1}= \omega_{n}^2+ \bk^2 + m_{\sigma,\pi}^2.
\end{equation}

\begin{figure}[h]
\epsfxsize= 15cm
\vspace{0.8cm}
\centerline{\epsffile{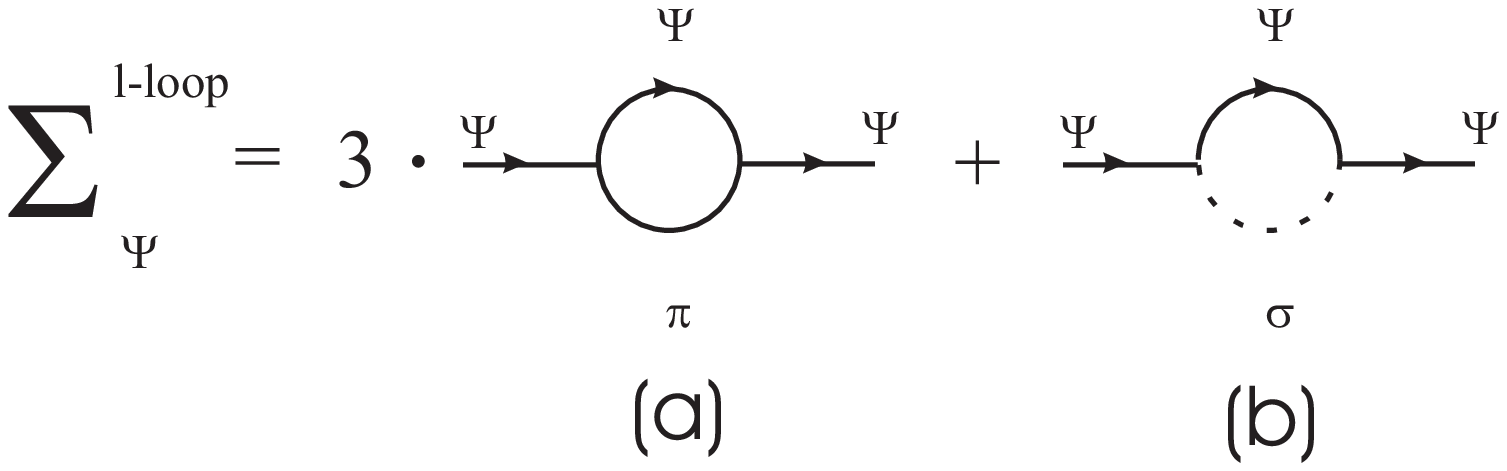} }
\vspace{0.8cm}
\protect\caption[]{The one-loop fermion self-energy. The pseudoscalar-boson contribution (a) and the scalar-boson contribution (b).}
\label{one}
\end{figure}

An evaluation of eq.(\ref{self2}) at zero three momentum gives $\Sigma(k_0, \left|\bk\right|=0)= (\gamma^0 \Sigma_0+\Sigma_s)_\sigma +3(\gamma^0 \Sigma_0(m_\sigma \leftrightarrow m_\pi)+\Sigma_s(m_\sigma \leftrightarrow m_\pi))_\sigma$ 
with the ``$\gamma^0$'' and ``scalar''- parts of the sigma contribution given respectively by 

\begin{eqnarray} 
\label{selfp1}
\Sigma_{0\sigma} = k_0  \frac{g^2}{2}\int_0^{\infty }\frac{dp p^2}{\pi^2} \frac {n_{\sigma}}{\omega_{\sigma}} 
\frac{-k_0^2+\omega_\sigma^2+\omega_\psi^2}{[ k_0^2-( \omega_\psi -\omega_\sigma)^2] [ k_0^2-( \omega_\psi +\omega_\sigma)^2]} 
\nonumber +\\
 k_0  \frac{g^2}{2}\int_0^{\infty }\frac{dp p^2}{\pi^2} 
\frac{n_{\psi}}{\omega_{\psi}} \frac{2\omega_\psi^2}{[ k_0^2-( \omega_\psi -\omega_\sigma)^2] [ k_0^2-( \omega_\psi +\omega_\sigma)^2]},
\end{eqnarray}

\begin{eqnarray} 
\label{Selfp2}
\Sigma_{s\sigma} =\frac{g^2}{2}\int_0^{\infty }\frac{dp p^2}{\pi^2} 
\frac{n_{\sigma}}{\omega_{\sigma}} \frac{m_\psi(-k_0^2-\omega_\sigma^2+\omega_\psi^2)}{[ k_0^2-( \omega_\psi -\omega_\sigma)^2] 
[ k_0^2-( \omega_\psi +\omega_\sigma)^2]} 
\nonumber +\\
\frac{g^2}{2}\int_0^{\infty }\frac{dp p^2}{\pi^2} \frac {n_{\psi}}{\omega_{\psi}} \frac{m_\psi(k_0^2-\omega_\sigma^2+\omega_\psi^2)}{[ k_0^2-( \omega_\psi -\omega_\sigma)^2] 
[ k_0^2-( \omega_\psi +\omega_\sigma)^2]},
\end{eqnarray}
where $\omega_{\sigma, \pi}^2 \equiv \bp^2 + m_{\sigma, \pi}^2$ and $\omega_{\psi}^2 \equiv \bp^2 + m_{\psi}^2$. The one-loop fermion self-energy is shown in Fig.\ref{one}.

\subsection{The Dispersion Relation Of Massless Fermions Interacting With Massless Bosons}

As a first approximation, in this subsection by considering the interaction of massless fermions with massless bosons we get an effective thermal fermion mass and also calculate the dispersion relation of the fermions.

The poles of the massless fermion propagator ($m_\psi=\Sigma_s=0$) gives the dispersion relation which occurs at the positive-energy root of

\begin{equation}
\label{root1}
[k_0(k)+\Sigma_0(k_0, k)]^2=\left| \vec k + \vec \Sigma(k_0,\vec k) \right|^2,
\end{equation}
where $k \equiv |\vec k|$.

For our intentions, it is sufficient to evaluate eq.(\ref{self2}) in the limit $k_0,k \ll T$, and consider the interaction of massless bosons and fermions in order to obtain an effective fermion thermal mass and dispersion relation. Thus,

\begin{equation}
\label{root2}
\Sigma_0(k_0, k)=-\frac{1}{8}g^2\frac{T^2}{k}\ln \left| \frac{k_0 + k}{ k_0 - k } \right|,
\end{equation}

\begin{equation}
\label{root3}
\vec \Sigma(k_0, k)=-\frac{1}{4}g^2\frac{T^2}{k^2}\left[\frac{k_0}{2k}\ln \left| \frac{k_0 + k}{ k_0 - k } \right|-1 \right]\vec k \equiv -\widetilde \Sigma \vec k,
\end{equation}
where we have defined 
$\frac{1}{4}g^2\frac{T^2}{k^2}\left[\frac{k_0}{2k}\ln \left| \frac{k_0 + k}{ k_0 - k } \right|-1 \right] \equiv \widetilde \Sigma$.

Some limits of expressions (\ref{root2}) and (\ref{root3}) are \cite{Kapusta2}:

\begin{eqnarray}
\label{li1}
\Sigma_0(k_0=0, k)=0, \\
\nonumber
\Sigma_0(k_0, k=0)=-\frac{g^2T^2}{4k_0}.
\end{eqnarray}

\begin{eqnarray}
\label{li2}
\vec \Sigma(k_0, k=0)=0, \\
\nonumber
\vec \Sigma(k_0=0, k)= \frac{g^2T^2}{4k} \hat k. 
\end{eqnarray}

Defining the fermion mass as the location of the pole in the limit $k=0$, we have $k_0+\Sigma_0(k_0,k \to 0)=0$, which implies

\begin{equation}
\label{root4}
M_\psi ^2=\frac{g^2T^2}{4}.
\end{equation}

From (\ref{root1}), we see that the fermion dispersion relation is given by 
$k_0+\Sigma_0=k(1+\widetilde \Sigma)$, that is

\begin{equation}
\label{root5}
k_0-\frac{1}{2}\frac{M_\psi ^2}{k}\ln \left| \frac{k_0 + k}{ k_0 - k } \right|
=k - \frac{M_\psi ^2}{k}\left[\frac{k_0}{2k}\ln \left| \frac{k_0 + k}{ k_0 - k } \right|-1 \right],
\end{equation}
which has the following well known form in the low momentum expansion\cite{Weldon1,Kapusta2}

\begin{equation}
\label{root6}
k_0(k)\equiv k_0(k)_+=M_\psi+\frac{1}{3}k+\frac{k^2}{3M_\psi}.
\end{equation}

This dispersion relation, $k_0(k)_+ \simeq M_\psi+\frac{1}{3}k$ for $k \ll M_\psi$, represents an ordinary fermion whose chirality is equal to its helicity. There is another dispersion relation, $ k_0(k)_- \simeq M_\psi-\frac{1}{3}k$, termed a plasmino \cite{Braaten3}, which describes a quasiparticle with chirality opposite to its helicity \cite{Braaten}.

\section{The Fermion Damping Rate}

Let us now proceed with the computation of the damping rate at rest ($k=0$). These calculations will be done considering in the first step $m_\psi=0$ and the 
dressed (effective) boson mass in the internal lines of the fermion self-energy. The effective fermion mass will be taken into account in the second step of the recalculation of the self-energy as the MSCR dictates \cite{Caldas1}. In the high temperature region, the bosons dressed masses (given by the MSCR) to be used in internal lines of the fermion self-energy read

\begin{equation}
\label{Mbosons}
m_\pi^2=m_\sigma^2 \equiv M_B ^2=\frac{g^2}{3}T^2.
\end{equation}

So, eq.(\ref{selfp1}) may be written as 

\begin{eqnarray} 
\label{selfp2}
\Sigma_0(k_0, \left|\bk\right|=0) =
-\frac{2g^2 \gamma^0 k_0}{\pi^2}\int_0^{\infty }dp p^2 \frac {n_{B}}{\omega_{B}} \frac{k_0^2-M_B^2-2p^2}{[ k_0^2-M_B^2]^2 -4k_0^2p^2} +\\
\nonumber
\frac{4g^2 \gamma^0 k_0}{\pi^2}\int_0^{\infty }dp p^4 \frac{n_{\psi}}{\omega_{\psi}} 
\frac{1}{[ k_0^2-M_B^2]^2 -4k_0^2p^2},
\nonumber
\end{eqnarray}
where $n_{B}$ and $n_{\psi}$ are the usual distribution functions 
for bosons and fermions given respectively by
\begin{equation}
\label{f1}
n_{B}(\omega_{B};T) = \frac{1}{ e^{ \beta \omega_{B}} - 1 } ,
\end{equation}

\begin{equation}
\label{f2}
n_{\psi}(\omega_{\psi};T) = \frac{1}{ e^{ \beta \omega_{\psi}} + 1 } .
\end{equation}
with $\omega_{B} \equiv \sqrt{\bp^2 + M_{B}^2}$ and $\omega_{\psi} \equiv \left |\bp \right|$.

It is worth to note that for $k_0 \approx M_B$ in eq.(\ref{selfp2}), it is easy to see that $\Sigma_0(k_0, \left|\bk\right|=0)$ reduces to the second line of eq.(\ref{li1}) which is a stable state without singularities and we have that there is no decay.

An explicit evaluation of eq.(\ref{selfp2}) in the pole of the correctec propagator (where the square of the $\gamma$ matrices is equal the unit matrix) furnishes 

\begin{eqnarray}
\label{root7}
\Sigma_0=-\frac{g^2}{4 \pi^2} \alpha \int_0 ^{\infty} dx \frac{n_B(\widetilde \omega_B)}{\widetilde \omega_B}\left[1+ \frac{\beta \alpha}{4}\left(\frac{1}{x-\beta\frac{\alpha}{2}}- \frac{1}{x+\beta\frac{\alpha}{2}}\right) \right] \\ 
-\frac{3g^2}{8 \pi^2} \alpha \int_0 ^{\infty} dx ~
n_\psi(\widetilde \omega_\psi) \left(\frac{1}{x-\beta\frac{\alpha}{2}}+ \frac{1}{x+\beta\frac{\alpha}{2}}\right) 
-\frac{3g^2T^2}{\pi^2 \alpha} \int_0^\infty dx ~ x^2 \left(\frac{n_B(\widetilde \omega_B)}{\widetilde \omega_B} + \frac{n_\psi(\widetilde \omega_\psi)}{\widetilde \omega_\psi} \right),
\nonumber
\end{eqnarray}
with the definitions $\alpha \equiv k_0 - \frac{M_B ^2}{k_0}$, $\widetilde \omega_B \equiv \sqrt{\beta^2 \bp^2 + \beta^2 M_{B}^2}$, $\widetilde \omega_{\psi} \equiv \left |\beta \bp \right|$ and $\beta p \equiv x$. The interesting physics happens when $k_0 > M_B$. Otherwise (if $k_0 < M_B$) one would get imaginary (forbidden) frequency.

The expression for $\Sigma_0$ in (\ref{root7}) has singularities, and now we 
adopt the prescription $\alpha = \omega - i \gamma$, since in general $k_0$ is 
complex, where $\omega$ is the real frequency and $\gamma$ is a real constant. With this assumption for $\alpha$, eq.(\ref{root7}) is expressed as

\begin{eqnarray}
\label{root8}
\Sigma_0=-\frac{g^2}{4 \pi^2} \alpha \int_0 ^{\infty} dx \frac{n_B(\widetilde \omega_B)}{\widetilde \omega_B}\left[1+ \frac{\alpha}{2}\left(\frac{\frac{2x}{\beta}- \omega}{\left(\frac{2x}{\beta}-\omega \right)^2+\gamma^2}- \frac{\frac{2x}{\beta}+ \omega}{\left(\frac{2x}{\beta}+\omega \right)^2+\gamma^2} \right) \right] +\\ 
\nonumber
i\frac{g^2}{8 \pi^2} \alpha^2 \int_0 ^{\infty} dx \frac{n_B(\widetilde \omega_B)}{\widetilde \omega_B}\left[\frac{\gamma}{\left(\frac{2x}{\beta}-\omega \right)^2+\gamma^2}+ \frac{\gamma}{\left(\frac{2x}{\beta}+\omega \right)^2+\gamma^2} \right]+\\
\nonumber
-\frac{3g^2}{4 \pi^2} \frac{\alpha}{\beta} \int_0 ^{\infty} dx 
n_\psi(\widetilde \omega_\psi) \left[\frac{\frac{2x}{\beta}- \omega}{\left(\frac{2x}{\beta}-\omega \right)^2+\gamma^2} + \frac{\frac{2x}{\beta}+ \omega}{\left(\frac{2x}{\beta}+\omega \right)^2+\gamma^2} \right] +\\
\nonumber
-i\frac{3g^2}{4 \pi^2} \frac{\alpha}{\beta} \int_0 ^{\infty} dx 
n_\psi(\widetilde \omega_\psi) \left[-\frac{\gamma}{\left(\frac{2x}{\beta}-\omega \right)^2+\gamma^2}+ \frac{\gamma}{\left(\frac{2x}{\beta}+\omega \right)^2+\gamma^2} \right] +\\
\nonumber
-\frac{3g^2T^2}{\pi^2 \alpha} \int_0^\infty dx x^2 \frac{n_B(\widetilde \omega_B)}{\widetilde \omega_B}
-\frac{3g^2T^2}{\pi^2 \alpha} \int_0^\infty dx x^2 \frac{n_\psi(\widetilde \omega_\psi)}{\widetilde \omega_\psi}.
\nonumber
\end{eqnarray}

Now making use of the definition of the delta function

\begin{equation}
\delta(y)= \stackrel{lim}{\epsilon\rightarrow0}  \frac{1}{\pi} \frac{\epsilon}{y^2+\epsilon^2},
\end{equation}
and the definitions

\begin{equation}
F(x,\omega) \equiv \frac{\frac{2x}{\beta}- \omega}{\left(\frac{2x}{\beta}-\omega \right)^2+\gamma^2}- \frac{\frac{2x}{\beta}+ \omega}{\left(\frac{2x}{\beta}+\omega \right)^2+\gamma^2},
\end{equation}
and

\begin{equation}
G(x,\omega) \equiv \frac{\frac{2x}{\beta}- \omega}{\left(\frac{2x}{\beta}-\omega \right)^2+\gamma^2}+ \frac{\frac{2x}{\beta}+ \omega}{\left(\frac{2x}{\beta}+\omega \right)^2+\gamma^2},
\end{equation}
we get

\begin{eqnarray}
\label{root9}
\Sigma_0=-\frac{g^2}{4 \pi^2} \alpha \int_0 ^{\infty} dx \frac{n_B(\widetilde \omega_B)}{\widetilde \omega_B}\left[1+ \frac{\alpha}{2}F(x,\omega) \right] + 
i\frac{g^2}{8 \pi} \left[\frac{\omega^2}{\sqrt{\frac{\omega^2}{4}+M_B^2}}\frac{1}{e^{\beta\sqrt{\frac{\omega^2}{4}+M_B^2}}-1} \right]+\\
\nonumber
-\frac{3g^2}{4 \pi^2} \frac{\alpha}{\beta} \int_0 ^{\infty} dx 
n_\psi(\widetilde \omega_\psi) G(x,\omega)+
i\frac{3g^2}{4 \pi} \frac{|\omega|}{e^{\frac{\beta|\omega|}{2}}+1} +\\
\nonumber
-\frac{3g^2T^2}{\pi^2 \alpha} \int_0^\infty dx x^2 \frac{n_B(\widetilde \omega_B)}{\widetilde \omega_B}
-\frac{3g^2T^2}{\pi^2 \alpha} \int_0^\infty dx x^2 \frac{n_\psi(\widetilde \omega_\psi)}{\widetilde \omega_\psi}.
\nonumber
\end{eqnarray}

Here we use some results from high temperature expansion of one-loop integrals derived by Dolan and Jackiw in \cite{Dolan}: 

\begin{equation}
\label{root10}
\int_0 ^{\infty} dx \frac{n_B(\widetilde \omega_B)}{\widetilde \omega_B}= \frac{\pi T}{2M_B} +\frac{1}{2}\ln \left(\frac{M_B}{4 \pi T} \right)+O \left( \frac{M_B^2}{T^2} \right)
\simeq \frac{\pi \sqrt{3}}{2 g},
\end{equation}
where in eq.(\ref{root10}) we have used from (\ref{Mbosons}) that $M_B=\frac{gT}{\sqrt{3}}$. 
This means that the first term in the r.h.s. of eq.(\ref{root9}) 
is $\frac{g \sqrt{3}}{8 \pi} \ll 1$. On the other hand, 
for the last two terms in the r.h.s. of this same expression, we have

\begin{equation}
\label{root11}
\frac{T^2}{\pi^2} \int_0^\infty dx x^2 \frac{n_B(\widetilde \omega_B)}{\widetilde \omega_B}
+\frac{T^2}{\pi^2} \int_0^\infty dx x^2 \frac{n_\psi(\widetilde \omega_\psi)}{\widetilde \omega_\psi}=\left( \frac{T^2}{6}-\frac{M_B T}{2 \pi}
+ O(g^2) \right)+ \left( \frac{T^2}{12} \right) \to \frac{T^2}{4}.
\end{equation}

The parts involving $F(x,\omega)$ and $G(x,\omega)$ are less important contributions in comparison to the dominant term that is proportional to $\frac{g^2T^2}{\omega}$, mainly for small $\omega$. So, in a first glance one can neglect them. Putting these results in eq.(\ref{root9}) and assuming ($\gamma \ll \omega$), the leading term of the real frequency can be written as

\begin{equation}
\label{root12}
\omega_1 ^2=\frac{9g^2T^2}{4} \equiv M_{\psi,1} ^2.
\end{equation}

The next step is the recalculation of the self-energy of the fermion to get the second order corrected fermion mass from the pole location

\begin{equation}
\label{pole}
k_{0,2}-M_{\psi,1}+\Sigma(M_B,M_{\psi,1},k=0)=0,
\end{equation}
where from this equation we will identify $(Re k_{0,2})^2=\omega_2^2 \equiv M_{\psi,2}^2$ and $Im k_{0,2}=\gamma$. With this resummation procedure it is possible to take into account the induced by the thermal medium fermion mass. At each recalculation we take into account more infinity subsets of graphs which result is shown in the mass. Now we are able to find out the damping 
rate of the fermion, which is defined by the imaginary part of the self-energy on-shell \cite{Braaten}.

From eq.(\ref{self2}) the self-energy $\Sigma(M_B,M_{\psi,1},k=0)=\Sigma_s(M_B,M_{\psi,1},k=0)+\Sigma_0(M_B,M_{\psi,1},k=0)$ in eq.(\ref{pole}) to be evaluated now is

\begin{eqnarray} 
\label{pole1}
\Sigma=-2 k_0 g^2 \int_0^{\infty }\frac{dp p^2}{\pi^2} \frac {n_{B}}{\omega_{B}} 
\frac{k_0^2-(M_B ^2+M_{\psi,1}^2+2p^2)}{[ k_0^2-(M_B ^2 + M_{\psi,1}^2)]^2 -4k_0^2p^2-4M_B ^2M_{\psi,1}^2} 
\nonumber +\\
4 k_0 g^2 \int_0^{\infty }\frac{dp p^2}{\pi^2} 
\frac{n_{\psi,1}}{\omega_{\psi,1}} \frac{M_{\psi,1}^2+p^2} {[ k_0^2-(M_B ^2 + M_{\psi,1}^2)]^2 -4k_0^2p^2-4M_B ^2M_{\psi,1}^2} 
\nonumber +\\
2 g^2 \int_0^{\infty }\frac{dp p^2}{\pi^2} 
\frac{n_{B}}{\omega_{B}} \frac{M_{\psi,1}(-k_0^2-M_B ^2 + M_{\psi,1}^2)}{[ k_0^2-(M_B ^2 + M_{\psi,1}^2)]^2 -4k_0^2p^2-4M_B ^2M_{\psi,1}^2} 
\nonumber +\\
2 g^2 \int_0^{\infty }\frac{dp p^2}{\pi^2} \frac {n_{\psi,1}}{\omega_{\psi,1}} \frac{M_{\psi,1}(k_0^2-M_B ^2 + M_{\psi,1}^2)}{[ k_0^2-(M_B ^2 + M_{\psi,1}^2)]^2 -4k_0^2p^2-4M_B ^2M_{\psi,1}^2}
\end{eqnarray}
where the last term ($4M_B ^2M_{\psi,1}^2$) in the denominators in the r.h.s. of eq.(\ref{pole1}) will be neglected, since it is of $O(g^4)$.

Doing again the calculations necessary to get the real and imaginary parts of the poles, where now $\alpha \equiv k_0 - \frac{M_B ^2+M_{\psi,1}^2}{k_0}$, $\widetilde \omega_B \equiv \sqrt{\beta^2 \bp^2 + \beta^2 M_{B}^2}$ and $\widetilde \omega_{\psi,1} \equiv \sqrt{\beta^2 \bp^2 + \beta^2 M_{\psi,1}^2}$, 
one finds

\begin{equation}
\label{root15}
\omega_2 ^2=M_{\psi,1} ^2 \left(1-\frac{2g}{\pi \sqrt{3}} \right) \equiv M_{\psi,2} ^2,
\end{equation}

\begin{equation}
\label{root13}
\gamma= \frac{3 g^2}{8 \pi} \left[\frac{\omega_2^2}{\sqrt{\frac{\omega_2^2}{4}+
M_B^2}}\frac{1}{e^{\beta\sqrt{\frac{\omega_2^2}{4}+M_B^2}}-1} \right]+ \frac{9 g^2}{8 \pi} \left[\frac{\omega_2^2}{\sqrt{\frac{\omega_2^2}{4}+
M_{\psi,1}^2}}\frac{1}{e^{\beta\sqrt{\frac{\omega_2^2}{4}+ M_{\psi,1}^2}}+1} \right].
\end{equation}

It is important to note that eq.(\ref{root15}) allows us to define a critical value of $g$ \cite{Bellac}, that is $g_{cr} \simeq 2.7$. This value agrees with the one obtained in ref \cite{Braaten}. As pointed out by M. Le Bellac in ref. \cite{Bellac}, collective excitations should disappear at the critical temperature, althoug we do not expect pertubation theory is still valid for such large values of $g$. Equations (\ref{root15}) and (\ref{root13}) has the following interpretation: The leading order of the real frequency $\omega_2$ is of order $gT$, in concordance with (\ref{root4}). The damping rate is proportional to the probability $n_\psi(\frac{1}{2}\omega)$ of having a fermion with energy $\frac{1}{2}\omega$ and a probability $n_B(\frac{1}{2}\omega)$ of having a boson with energy $\frac{1}{2}\omega$. These probabilities are weighted by numerical factors and the available phase space $\omega$ \cite{Kapusta}. The distribution functions can be expanded for low energies, $n_B(\frac{1}{2}\omega) \simeq T/ \omega$ and $n_\psi(\frac{1}{2}\omega) \simeq 1/2$ and the damping rate (up to order $g^3$) reduces to

\begin{equation}
\label{root14}
\gamma \simeq \frac{9 g^2 T}{16 \pi}+ \frac{27g^3T}{32 \pi}.
\end{equation}

\section{Concluding Remarks}
\label{conc} 
In this paper, we have considered the fermion boson interaction at finite temperature. First, we have calculated the self-energy of the fermions due the interaction with scalar-bosons and pseudoscalar-bosons in the framework of the linear sigma model. 

Next, we have calculated the fermion dispersion relation in the limit 
$k_0,k \ll T$ of massless fermions interacting with massless bosons and some of its limits. Also, we have obtained the thermal fermion mass which is of order $gT$.

Finally, we have computed the frequency and the damping rate of the fermion at rest, considering the dressed fermion and boson masses in the internal lines of the fermion self-energy rather than the zero mass parameter of the Lagrangian. The damping rate of the fermion was found to be of order $g^2T$ from the boson internal line of the fermion self-energy plus a part which is of order $g^3T$ from the fermion internal line of the self-energy. This is a remarkable result since it shows the signature of the alternative MSCR method (up to second order of the non-perturbative correction) in the desired effective mass and damping rate. One important feature of this calculation (up to this order) is that we have gotten this result algebraically in a clear manner differently from the cases where the results are reached numerically. 

The calculation of the fermion damping rate at rest constitutes another simple but instructive application of the MSCR method.

\section*{Acknowledgements}
The author thanks the hospitality given by the 
Nuclear Theory group during his visit at the University of Minnesota were 
this work was initiated. He is gratefully indebted to Professors Joe Kapusta 
and Paul Ellis for various helpful discussions. Also, the author would like 
to thank Professor Paul Ellis for a critical reading of the manuscript.

\newpage

%%%%%%%%%%%%%%%%%%%%%%%%%%%%%%%%%%%%%%%%%%%%%%%%%%%%%%%%%%%%%%%%%

\newpage

\begin{references} 
\let\ul=\underbar
\def\AP#1{{Ann. Phys. }{\bf #1}}
\def\PRP#1{{Phys. Rep. }{\bf #1}}
\def\APP#1{{Act. Phys. Pol. }{\bf #1}}
\def\PTP#1{{Prog. Theor. Phys. }{\bf #1}}
\def\PR#1{{Phys. Rev. }{\bf #1}}
\def\PRD#1{{Phys. Rev. }{\bf D #1}}
\def\PRC#1{{Phys. Rev. }{\bf C #1}}
\def\PRL#1{{Phys. Rev. Lett. }{\bf #1}}
\def\PL#1{{Phys. Lett. }{\bf #1}}
\def\RMP#1{{Rev. Mod. Phys.}{\bf #1}}
\def\NP#1{{Nucl. Phys. }{\bf #1}}
\def\ZP#1{{Z. Phys. }{\bf #1}}
\def\NC#1{{Nuovo Cimento }{\bf #1}}
\def\SJNP#1{{Sov. J. Nucl. Phys. }{\bf #1}} 

\bibitem{Weldon1} H.A. Weldon, \PRD{26}, 2789 (1982).
\bibitem{Kapusta1} G. Gatoff and J. Kapusta, \PRD{41}, 611 (1990).
\bibitem{Braaten} E. Braaten and R.D. Pisarski, \PRD{46}, 1829 (1992).
\bibitem{Letessier} J. Letessier, J. Rafelski and A. Tomsi, \PL{B 323}, 393 (1990).
\bibitem{Braaten1} E. Braaten, R.D. Pisarski and T.C. Yuan, \PRL{66}, 2183 (1991).
\bibitem{Thoma1} M.H. Thoma and C.T. Traxler, \PRD{56}, 198 (1997).
\bibitem{Daniel1} D. Boyanovsky, H.J. de Veja, D.-S. Lee, Y.J. Ng 
and S.-Y. Wang, \PRD{59} 105001 (1999), hep-ph/9810393.
\bibitem{Weldon2} H.A. Weldon, \PRD{61}, 036003 (2000).
\bibitem{Rebhan1} Anton Rebhan, \PRD{46}, 4779 (1992).
\bibitem{Rebhan2} R. Kobes, G. Kunstatter and A. Rebhan, \PRL{64}, 2992 (1990); \NP{B355}, 1 (1991).
\bibitem{Braaten0} E. Braaten and R.D. Pisarski, \NP{B337}, 569 (1990).
\bibitem{Braaten2} E. Braaten and R.D. Pisarski, \NP{B337}, 369 (1990); E. Braaten and R. D. Pisarski, \PRL{64}, 1338 (1990).
\bibitem{Thoma} M. H. Thoma, \ZP{C66}, 491 (1995).
\bibitem{Caldas1} H.C.G. Caldas, A.L. Mota and M.C. Nemes, \PRD{63}, 56011 (2001), hep-ph/0005180.
\bibitem{Caldas1-2} H.C.G. Caldas, to appear in Phys. Rev. D, hep-th/0111194.
\bibitem{Weinberg} S. Weinberg, \PRD{\bf 9}, 3357 (1974).
\bibitem{Dolan} L. Dolan and R. Jackiw, Phys. Rev. D {\bf 9}, 3320 (1974).
\bibitem{Caldas2} H.C.G. Caldas, work in progress.
\bibitem{Gell-Mann} M. Gell-Mann, and M. Levy, \NC{16}, 705 (1960).
\bibitem{Bellac} Michel Le Bellac, {\it Thermal Field Theory}
(Cambridge University Press, Cambridge, 1996).
\bibitem{Kapusta} J. Kapusta, {\it Finite-Temperature Field Theory}
(Cambridge University Press, Cambridge, 1989).
\bibitem{Kapusta2} J. Kapusta, CERN preprint, 1982 (unpublished).
\bibitem{Braaten3} E. Braaten, Astrophys. J. {\bf 392}, 70 (1992).


\end{references}
\end{document}